# Multiplicity predictions for Pb-Pb collisions at $\sqrt{s}_{NN}$=5.02 TeV based on the MC Modified Glauber model

**A. Seryakov, G. Feofilov**
St. Petersburg State University, Russia

E-mail: seryakov@yahoo.com

**Abstract.** We present predictions of the mean multiplicity yields for different centrality classes of Pb-Pb collisions at $\sqrt{s}_{NN}$=5.02 TeV at the LHC, obtained using the Monte-Carlo Modified Glauber model (MGM) [1]. Contrary to generally used standard Glauber calculations, the MGM takes into account the energy conservation in the soft particles production. The only one efficient model parameter was tuned and fixed previously (k=0.35) by fitting the available data on multiplicity for all colliding systems up to RHIC energies. The MGM has shown already a good predictive capability for the yields at $\sqrt{s}_{NN}$=2.76 TeV, and it is used again within the same MC code for the calculations at $\sqrt{s}_{NN}$=5.02 TeV.

## 1. Monte-Carlo Modified Glauber model

The main idea of the Modified Glauber model (MGM [1]) is to introduce the energy conservation law into the process of particle production. Contrary to the widely used standard Glauber model (SGM), a nucleon in the MGM loses in nucleon-nucleon collision the fixed part (k) of its momentum in the center of mass system. This single efficient model parameter k was tuned and fixed (k=0.35) by fitting the available data on mean multiplicity in the most central events for all heavy ion colliding systems up to RHIC energies. Therefore, with the account of the relevant energy loss, the following nucleon-nucleon collision is treated at the decreased collision energy and with the different value of the inelastic cross-section. Some other effects were also added later [2] [3]. We present briefly below the current status of the MGM in addition to the Glauber model concepts.

Standard Glauber model:
- Nucleus-nucleus collision is a superposition of independent nucleon-nucleon collisions.
- Nucleons have linear trajectories.
- Distribution of nucleons inside the colliding nuclei are taken from the experimental data [4].
- Probability of nucleon-nucleon interaction is taken from the experimental pp inelastic cross-section [5] with an assumption that a nucleon is a black disk.

Modified Glauber model:
- Nucleon in each nucleon-nucleon collision loses the fixed part of the momentum in the center of mass system, so that $p' = kp$, where $p$ is the momentum before collision, $p'$ - after the collision. Here $k$ is the momentum loss coefficient, which is considered to be **the same** for all energies.
- Multiplicity of nucleus-nucleus collision is a sum of multiplicities of all individual nucleon-nucleon collisions, so that $N_{ch}^{AA} = \sum_{Ncoll} N_{ch_i}$. Note that this is not equal to $Ncoll \cdot < N_{ch}^{pp} >$, where $<N_{ch}^{pp}>$ is the mean multiplicity total yield in *pp* collisions at the given energy. This is mainly due to the fact that each nucleon-nucleon collisions take place at the energy defined with the account of the energy loss during the previous nucleus-nucleus collision. Besides this, the account of charge conservation is found to be noticeable.

- Multiplicity of the nucleon-nucleon collision is taken from the experimental data [6] with the correction of charge conservation: $N_{ch}(\sqrt{s}) = N_{ch}^{pp}(\sqrt{s}) - 2 + nucleons\_charges$. It means that the charge of particular nucleon is not counted twice in the collision.
- Secondary collisions of nucleons are also taken into account. It is the nucleon-nucleon collisions of nucleons which belong to the same nucleus and are produced due to the kinematical effects when one of the nucleons has lost a significant part of his momentum after several inelastic collisions (and even changed the direction of motion).

The momentum loss parameter $k = 0.35$ was obtained in [1] by fitting multiplicity distributions only for central collisions for energies up to $\sqrt{s} = 200$ GeV.

## 2. Results

The comparisons of the MGM predictions to the already available Pb-Pb and Au-Au data on total charge particle yield per the number of nucleon-participants are shown in the fig.2.

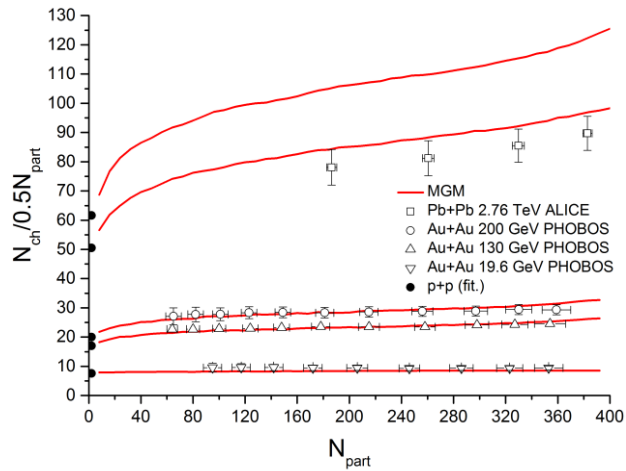

**Figure 2.** Multiplicity for charged particles to the pair of nucleon-participants versus centrality of nucleus-nucleus collisions, defined. The open points are ALICE data [7] at 2.76 TeV and RHIC data at three collision energies [6]. The protons points were taken from the fitting of pp data [6]. The red curves are calculated in MGM with k = 0.35 at relevant collision energies (no fits). The upper red curve is the prediction for Pb-Pb at $\sqrt{s}_{NN} = 5.02$ TeV.

**References**
[1]  G. Feofilov, A. Ivanov, Journal of Physics G CS, 5, (2005) 230-237
[2]  G. Feofilov, A. Seryakov, PoS(Baldin ISHEPP XXII)082
[3]  T. Drozhzhova, G. Feofilov, V. Kovalenko, A. Seryakov, PoS(QFTHEP 2013)053
[4]  H. Vries, C. Jager, C. Vries, Atomic data and nuclear data tables 36, 495-536 (1987)
[5]  TOTEM Coll, arXiv:1204.5689 [hep-ex], 2012.
[6]  PHOBOS Coll., Phys.Rev.C83:024913,2011.
[7]  ALICE Coll., CERN-PH-EP-2013-045